\documentclass[apj,numberedappendix]{emulateapj}
\usepackage{times}
\usepackage[normalem]{ulem}
\usepackage{epsfig}
\usepackage{natbib}
\usepackage{amsfonts}
\usepackage{amsmath}
\usepackage{multirow}
\usepackage{enumerate}
\usepackage{verbatim}
\usepackage{chngcntr}

\usepackage[usenames]{color}
\usepackage[plainpages=false, colorlinks=true, anchorcolor=blue, link
color=blue, citecolor=blue, bookmarks=false]{hyperref}
\newcommand{\be}{\begin{eqnarray}}
\newcommand{\ee}{\end{eqnarray}}

\newcommand{\lp}{\left(}
\newcommand{\rp}{\right)}
\newcommand{\lb}{\left[}
\newcommand{\rb}{\right]}

\newcommand{\slugcom}{Accepted for publication in The Astrophysical Journal}
\slugcomment{\slugcom}

\begin{document}

\normalsize

% -----------------------------------------------------------
% -----------------------------------------------------------

\title{The Dispersion and Rotation Measure of Supernova Remnants and\\ Magnetized Stellar Winds: Application to Fast Radio Bursts}

\author{Anthony L. Piro\altaffilmark{1}}
\author{B. M. Gaensler\altaffilmark{2,3}}

\altaffiltext{1}{The Observatories of the Carnegie Institution for Science, 813 Santa Barbara St., Pasadena, CA 91101, USA; piro@carnegiescience.edu}

\altaffiltext{2}{Dunlap Institute for Astronomy and Astrophysics, University of Toronto, Toronto, ON M5S 3H4, Canada; bgaensler@dunlap.utoronto.ca}

\altaffiltext{3}{Department of Astronomy and Astrophysics, University of Toronto, Toronto, ON M5S 3H4, Canada}

\begin{abstract}
Recent studies of fast radio bursts (FRBs) have led to many theories associating them with young neutron stars. If this is the case, then the presence of supernova ejecta and stellar winds provide a changing dispersion measure (DM) and rotation measure (RM) that can potentially be probes of the environments of FRB progenitors. Here we summarize the scalings for the DM and RM in the cases of a constant density ambient medium and of a progenitor stellar wind. Since the amount of ionized material is controlled by the dynamics of the reverse shock, we find the DM changes more slowly than in previous simpler work, which simply assumed a constant ionization fraction. Furthermore, the DM can be constant or even increasing as the supernova remnant sweeps up material, arguing that a young neutron star hypothesis for FRBs is not ruled out if the DM is not decreasing over repeated bursts. The combined DM and RM measurements for the repeating FRB 121102 are consistent with supernova ejecta with an age of $\sim10^2-10^3\,{\rm yrs}$ expanding into a high density ($\sim100\,{\rm cm^{-3}}$) interstellar medium. This naturally explains its relatively constant DM over many years as well. Other FRBs with much lower RMs may indicate that they are especially young supernovae in wind environments or that their DMs are largely from the intergalactic medium. We therefore caution about inferring magnetic fields from simply by dividing an RM by DM, because these quantities could originate from distinct regions along the path an FRB propagates.
\end{abstract}

	
\section{Introduction}

Fast radio bursts (FRBs) are a class of transients characterized by millisecond flashes of radio radiation \citep{Lorimer07,Keane12,Thornton13,Ravi15}. Their large dispersion measures (DMs) and Faraday rotation measures (RMs) imply that they likely occur at cosmological distances and/or in extreme density environments \citep[see discussions by][and references therein]{Kulkarni14,Luan14,Lyubarsky14,Katz16}. An important constraint on their origin is that they appear to be very common, with an inferred rate of $\sim10^3-10^4$ FRBs on the sky per day \citep[e.g.,][]{Rane16,VanderWiel16,Bhandari17}. Nevertheless, there have been no astrophysical objects definitively connected to FRBs, leaving their DMs and RMs as vital probes as to their mechanisms, progenitors and environments.

There are multiple possible contributions to the DM and RM of an FRB. These include the disk of the Milky Way \citep{Oppermann2012,Yao16}, the Milky Way's halo \citep{Dolag15}, the intervening intergalactic medium \citep{McQuinn2014,Akahori2016}, the corresponding disk and halo of an FRB's host galaxy \citep{Xu15,Tendulkar17}, and the FRB's immediate local environment \citep{Connor16,Lyutikov16,Piro16,Yang17b,Michilli18}.

There are a variety of arguments that FRBs are produced by young neutron stars \citep{Popov10,Waxman17,Nicholl17}. Since neutron stars are formed in core-collapse supernova (SN) explosions, FRB signals should thus pass through the expanding shell of a young supernova remnant (SNR), which should make a corresponding contribution to the FRB's DM and RM. An important conclusion emphasized by \citet{Piro16} is that even though the SNR may not dominate the {\it total} DM or RM, it should dominate the {\it change} in the DM or RM seen with time, as might be discernible over a time scale of several years. In this way, the environment and ultimately the source of the FRB may be better understood for a repeating FRB \citep{Spitler16,Piro17}. These contributions should show a secular decrease at early times as the SNR expands ballistically \citep{Piro16,Katz16c,Murase16,Metzger17,Yang17a,Yang17b}, but the DM from the SNR should be constant or even increasing with time once the SNR has swept up an amount of material similar to the ejecta mass \citep{Piro16,Yang17b}.

%In the work of \cite{Piro16}, it was found that by the time that a SNR is optically thin to FRB signals (at an SNR age of $\sim$50--500~years), the DM and RM should have dropped to less than $\sim$100~pc~cm$^{-3}$ and $\sim$100~rad~m$^{-2}$, respectively. Even though these SNR contributions might not dominate the {\it total} DM or RM at any given time, they should dominate the {\it change} in the DM or RM seen with time, as might be discernible over a time scale of several years for a repeating FRB \citep[see also][]{Piro17,Yang17b}. Furthermore, \cite{Piro16} mostly focused on normal, massive SNRs, but recent work has argued that the SN events that give rise to FRB generators may be especially mass-stripped or energetic \citep{Piro17,Metzger17}. Thus it is important to revisit these arguments for such a parameter space.

Although \citet{Piro16} and \citet{Yang17b} provide the most complete description of the SNR impact thus far, important details still remain to be explored. First, the dynamics of the reverse shock is critical for understanding the amount and geometry of the ionized material that can disperse the FRB. Although this was included by \citet{Piro16}, the difference this introduces to the scalings with time was not sufficiently highlighted, nor was this included in subsequent works (which typically assume a constant ionized fraction).

Another important issue is that core-collapse progenitors are massive stars that will have strong, magnetized, winds \citep{icb98,ud02}. As this wind is swept up by the expanding SNR \citep{Chevalier82,Chevalier03,Harvey-Smith10}, it can be an important additional contribution to the DM and RM of an FRB. Furthermore, the decreasing density profile with radius of a wind can impact the dynamics of an SNR differently than a constant density ISM as used by \citet{Piro16} and \citet{Yang17b}.

Motivated by these issues, we investigate in further detail the DM and RM seen for an FRB and their time evolution due to an SNR and its environment. In Section~\ref{sec:constant}, we consider the contributions of the SNR and a constant density interstellar medium (ISM), from the blast wave through Sedov-Taylor phases of evolution. In Section~\ref{sec:wind}, we instead consider a magnetized stellar wind environment and highlight the distinct DM and RM evolution. We discuss the implications of these results for observations of FRBs in Section~\ref{sec:discussion}, and conclude with a summary of our work in Section~\ref{sec:conclusions}.

\section{Constant Density ISM}
\label{sec:constant}

\begin{figure}
\epsscale{1.15}
\plotone{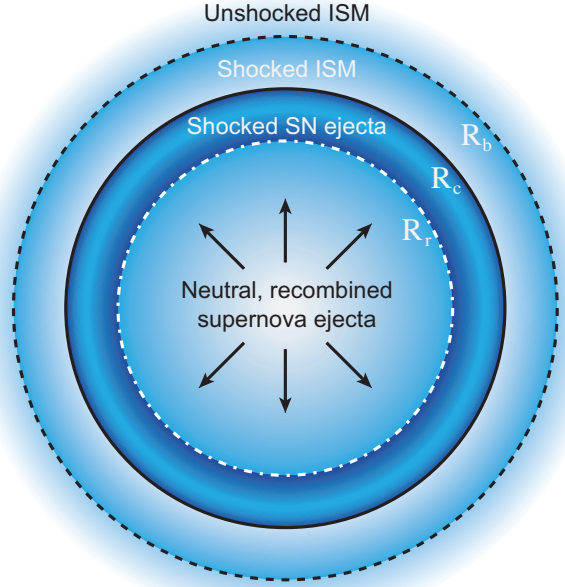}
  \caption{{Schematic showing the main regions of focus for the SNR. The key radii are at the positions of the reverse shock $R_r$ (white dot-dashed line), contact discontinuity $R_c$ (black solid line), and the forward shock or blast wave $R_b$ (black dashed line). The main ionized regions, which can contribute to the DM and RM of an FRB, sit between $R_r$ and $R_b$. These are composed of (1) the shocked SN ejecta (between the radii of $R_r$ to $R_c$) and (2) the shocked ISM (between the radii of $R_c$ to $R_b$).}}
  \label{fig:diagram}
\end{figure}

We first describe the evolution of an SN expanding into a constant density ISM. The mass distribution can be roughly divided into four regions that are summarized in Figure \ref{fig:diagram}. These are, in order of increasing radius: (1) neutral, recombined SN ejecta, (2) shocked SN ejecta, (3) shocked ISM material, and (4) unshocked ISM. These are separated by three key radii: (1) the reverse shock, at radius $R_r$, (2) the contact discontinuity between the SN ejecta and ISM, at $R_c$, and (3) the forward shock or blastwave radius, at $R_b$. To understand FRBs propagating through this material from an embedded central neutron star, we focus on the two shocked, ionized regions that provide sufficient free electrons to significantly disperse the FRB signal {(the region between $R_r$ and $R_b$ in Figure \ref{fig:diagram}).} In particular, in this work we make a better distinction between $R_c$ and $R_b$ in comparison to \citet{Piro16}. {An additional source of ionized material comes from the pulsar wind nebula located near the center of the SNR. Even though the amount of ionizing emission can be especially strong in the case of a highly magnetized neutron star \citep{Metzger17},  it still is a small contribution in comparison to the outer shocked material, and so we save a detailed study of this for future work.}

As the SN ejecta expand, they roughly evolve through two stages. This is summarized by the approximate analytic functions provided in Table \ref{table} (from the work of \citealp{McKee95}; see also the plotting of these functions in Figure \ref{fig:snr_sln}). First, the ejecta will be in an ``ejecta-dominated phase,'' for which the blastwave radius $R_b$ is moving at roughly constant velocity independent of the density of material surrounding the SN. This continues up until the time when the SN has swept up an amount of material approximately equal to the mass of the eject. This occurs on the Sedov-Taylor timescale\footnote{The exact numerical values for $t_{\rm ST}$ and $R_{\rm ST}$ are set by the continuity conditions at time $t_{\rm ST}$ as described by \citet{McKee95}. One can see analogous arguments used for an SNR expanding into a wind environment in Appendix~\ref{sec:appendix}.},
\be
	t_{\rm ST} = 210\,E_{51}^{-1/2} M_1^{5/6} n_0^{-1/3}\,{\rm yr}, 
\ee
where $E=10^{51}E_{51}\,{\rm erg}$ is the energy of the explosion, $M=M_1M_\odot$ is the mass of the SN ejecta, and $n_0$ (in units of cm$^{-3}$) is the number density of a uniform ambient ISM. Associated with this are the characteristic length scale,
\be
	R_{\rm ST} = 2.2M_1^{1/3}n_0^{-1/3}\,{\rm pc},
\ee
and velocity,
\be
	v_{\rm ST} = \frac{R_{\rm ST}}{t_{\rm ST}}
    = 1.0\times10^4E_{51}^{1/2} M_1^{-1/2}{\rm km\,s^{-1}}.
\ee
In the second stage, after a time $t_{\rm ST}$, the expansion of the ejecta slows as summarized in the right column of Table \ref{table}.

The velocities in Table \ref{table} refer to the velocities of the forward and reverse shocks. In particular, $\tilde{v}_r$ is in the rest frame of the unshocked ejecta just ahead of it, $\tilde{v}_r\equiv R_r/t-dR_r/dt$, rather than the rest frame reverse shock velocity $v_r\equiv dR_r/dt$. This is because it is the former quantity that is most relevant for estimating properties of the ejecta, such as the shock temperature and pressure. The contact discontinuity $R_c$ is estimated from the mass conservation condition,
\be
	\frac{4\pi}{3}R_b^3 n_0\approx\frac{4\pi}{3} (R_b^3-R_c^3)4n_0,
\ee
where we have used the compressibility of a $\gamma=5/3$ strong shock condition $(\gamma+1)/(\gamma-1)=4$. This then gives $R_c~\approx~(3/4)^{1/3}R_b$ as we use in Table \ref{table} for both $t<t_{\rm ST}$ and $t>t_{\rm ST}$. Such a relation is most accurate at early times, but gets increasingly poor at later times when the density is not constant across the reverse shocked region \citep[for example, see the study by][]{Tang17}. At least for the work here, this is a sufficient approximation, and we save a more detailed numerical treatment for future investigations.

The evolution of the SNR is summarized with the fiducial values of $M=1\,M_\odot$, $E=10^{51}\,{\rm erg}$, and $n_0=1\,{\rm cm^{-3}}$ in Figure \ref{fig:snr_revisit}. This shows how the SN blastwave radius roughly evolves as
\be
  R_b\propto\begin{cases}
    t, & t\lesssim t_{\rm ST}\\
    t^{2/5}, & t\gtrsim t_{\rm ST},
  \end{cases}
\ee
from the ejecta-dominated to Sedov-Taylor stages. The analytic expressions given in Table~\ref{table} allow us to follow the smooth evolution of the SNR between these limits. Figure~\ref{fig:snr_revisit} also shows how narrow the ionized regions are in radius, especially during the early phases.

Following the Sedov-Taylor stage, there is the ``snowplow stage'' when the SNR begins to radiatively cool appreciably. This roughly occurs at a time \citep{Draine11,Yang17b}
\be
	t_{\rm snow} \approx 4.9\times10^4 E_{51}^{0.22}n_0^{-0.55}\,{\rm yrs}.
\ee
We do not consider this stage in detail for this work, and thus our solutions and discussions are only applicable up until this timescale.

\subsection{Constant Density: Dispersion Measure}

For an FRB at redshift $z$, and assuming that the Milky Way component can be subtracted out, the remaining total DM is
\be
	{\rm DM}_{\rm tot}
    = \frac{{\rm DM}_{\rm local}}{1+z}
    +\frac{{\rm DM}_{\rm host}}{1+z}
    + {\rm DM}_{\rm IGM},
    \label{eq:dmtot}
\ee
where ${\rm DM}_{\rm host}$ is the contribution from the FRB host galaxy, ${\rm DM}_{\rm IGM}$ is the contribution from the intervening intergalactic medium (IGM), and
\be
	{\rm DM}_{\rm local} = {\rm DM}_{\rm SNR} + {\rm DM}_{\rm ISM},
\ee
is the local contribution from the shocked SN material and shocked ISM, respectively. In Section~\ref{sec:wind}, we consider a wind profile for the material around the SN instead, which also adds a contribution ${\rm DM}_w$ to ${\rm DM}_{\rm local}$. The IGM component can be approximated as \citep{Katz16c}
\be
	{\rm DM}_{\rm IGM} = \frac{n_{\rm IGM}c}{H_0}z,
\ee
where $H_0$ is Hubble's constant and $n_{\rm IGM}$ is the present-day density of the IGM ($n_{\rm IGM}=1.6\times10^{-7}\,{\rm cm^{-3}}$, assuming that the baryons are homogeneously distributed and ionized). A more detailed expression for this term is provided by \citet{Deng14}.

  \begin{deluxetable*}{lccc}
  \tablecolumns{10} \tablewidth{500pt}
 \tablecaption{Evolution for Constant Density ISM\tablenotemark{$a$}}
   \tablehead{  & Ejecta-Dominated Stage $(t<t_{\rm ST})$ & Sedov-Taylor Stage $(t>t_{\rm ST})$}
  \startdata
   Forward Shock &  $R_b/R_{\rm ST} = 1.37(t/t_{\rm ST})\lb 1+0.60(t/t_{\rm ST})^{3/2}\rb^{-2/3}$ & $R_b/R_{\rm ST}= \lb 1.56(t/t_{\rm ST})-0.56\rb^{2/5}$ \\
    & $v_b/v_{\rm ST} = 1.37\lb 1+0.60(t/t_{\rm ST})^{3/2} \rb^{-5/3}$  & $v_b/v_{\rm ST} = 0.63\lb 1.56(t/t_{\rm ST})-0.56\rb^{-3/5}$ \\
	Contact Discontinuity &  $R_c/R_{\rm ST} = 1.24(t/t_{\rm ST})\lb 1+0.60(t/t_{\rm ST})^{3/2}\rb^{-2/3}$ & $R_c/R_{\rm ST}= 0.91\lb 1.56(t/t_{\rm ST})-0.56\rb^{2/5}$ \\
    & $v_c/v_{\rm ST} = 1.24\lb 1+0.60(t/t_{\rm ST})^{3/2} \rb^{-5/3}$  & $v_c/v_{\rm ST} = 0.57\lb 1.56(t/t_{\rm ST})-0.56\rb^{-3/5}$ \\
   Reverse Shock & $R_r/R_{\rm ST} = 1.24(t/t_{\rm ST})\lb1+1.13(t/t_{\rm ST})^{3/2} \rb^{-2/3}$ & $R_r/R_{\rm ST} = (t/t_{\rm ST})\lb 0.78 -0.03(t/t_{\rm ST})-0.37\ln (t/t_{\rm ST}) \rb $\\
    & $\tilde{v}_r/v_{\rm ST} = 1.41(t/t_{\rm ST})^{3/2}\lb 1+1.13(t/t_{\rm ST})^{3/2} \rb^{-5/3}$ & $\tilde{v}_r/v_{\rm ST} = 0.37+0.03(t/t_{\rm ST})$\\
           DM (${\rm pc\,cm^{-3}}$)\tablenotemark{$b$} & $52.6 (\mu/\mu_e) E_{51}^{-1/4}M_1^{3/4}n_0^{1/2}t_{\rm yr}^{-1/2}$ & $1.8E_{51}^{1/5}n_0^{4/5}t_{1000\,\rm yr}^{2/5}$\\
       $|$RM$|$ (${\rm rad\,m^{-2}}$)\tablenotemark{$b,c$} & $ 1.8\times10^5(\mu^{3/2}/\mu_e) \epsilon_{-1}^{1/2} E_{51}^{1/4}  M_1^{1/4}n_0t_{\rm yr}^{-1/2}$ & $810 \epsilon_{-1}^{1/2}E_{51}^{2/5}  n_0^{11/10}t_{\rm 1000\,yr}^{-1/5}$
   \enddata
    \tablenotetext{$a$}{Relations from \citet{McKee95}. Note that $\tilde{v}_r$ is not the rest frame velocity of the reverse shock but rather the velocity in the frame of the unshocked ejecta just ahead of it, $\tilde{v}_r \equiv R_r/t - dR_r/dt$.}
        \tablenotetext{$b$}{These expressions for DM and RM are in the extreme limits of $t\ll t_{\rm ST}$ and $t\gg t_{\rm ST}$. For the more detailed evolution, one should consult \mbox{Figures \ref{fig:dm_constant} and \ref{fig:rm_constant}.}}
    \tablenotetext{$c$}{Upper limit since the parallel magnetic field could be smaller than the field assumed for this estimate.}
\label{table}
\end{deluxetable*}

For determining the DM that may be imprinted on an FRB by the SNR, we must consider each of the regions and the different stages of the evolution. For the SN ejecta, only the region from $R_r$ out to $R_c$ is ionized. Thus, integrating through the ionized material, the dispersion measure of the SNR is given by
\be
	{\rm DM}_{\rm SNR} =\int_{R_r}^{R_c} n_r dl
	\approx n_r(R_c-R_r),
    \label{eq:dm_snr}
\ee
where $n_r$ is the number density of electrons behind the reverse shock. This density is somewhat higher than the average density of the remnant, and can be determined by assuming pressure continuity across the contact discontinuity, which gives a reverse shock mass density of
\be
	\frac{\rho_r}{\mu m_p} \approx 4n_0 \left( \frac{v_b}{\tilde{v}_r} \right)^2,
    \label{eq:rho_r}
\ee
where $\mu$ is the mean molecular weight. The actual electron number density in the reverse shock region is
\be
	n_r = \rho_r/\mu_e m_p,
    \label{eq:ner}
\ee
where $\mu_e$ is the mean molecular weight per electron.

\begin{figure}
\epsscale{1.2}
\plotone{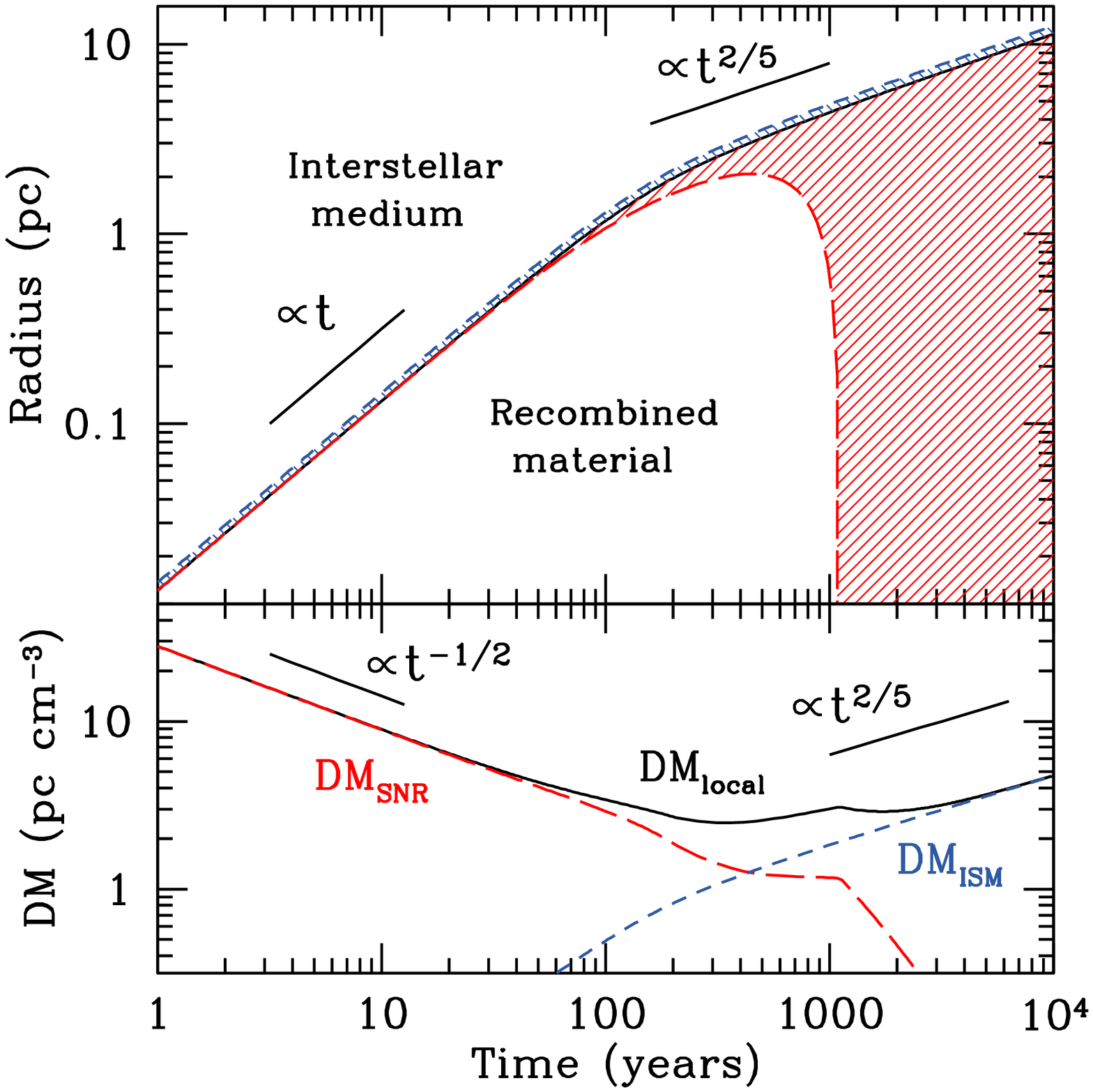}
\caption{Sample evolution of a SNR and the resulting DM for fiducial values $M=1\,M_\odot$ and $E=10^{51}\,{\rm erg}$, expanding into a neutral ($f=0$) uniform ISM of number density $n_0=1\,{\rm cm^{-3}}$; this combination corresponds to $t_{\rm ST} = 210$~yr. The top panel shows the evolution of the three key radii $R_r$ (red long-dashed line), $R_c$ (black solid line), and $R_b$ (blue short-dashed line). The red and blue shaded regions denote the shocked SN ejecta and ISM, respectively. The bottom panel shows the DM solution using Equations~(\ref{eq:dm_snr}) and (\ref{eq:dm_ism}). This evolves from being SN ejecta dominated during the blastwave stage (red long-dashed line) to ISM dominated during the Sedov-Taylor stage (blue short-dashed line).}
\label{fig:snr_revisit}
\epsscale{1.0}
\end{figure}

For the ISM contribution to the DM, assuming that this is mostly hydrogen dominated, it is
\be
	{\rm DM}_{\rm ISM} &=& 4n_0 (R_b-R_c) + fn_0(R_{\rm ISM}-R_b),
    \label{eq:dm_ism}
\ee
where the factor of $4$ is from compression of material at the forward shock. The term on the far righthand side corresponds to a possible contribution from ionized ISM material surrounding the SNR \citep{Yang17b}, where $R_{\rm ISM}$ is the extent of this region and $f$ is the ionized fraction. For the most part, we ignore this contribution when presenting ${\rm DM}_{\rm local}$, but we do discuss it further below since if may be important for the time-changing DM.

Using Equations (\ref{eq:dm_snr}) and (\ref{eq:dm_ism}) with the expressions for $R_r$, $R_c$, and $R_b$ from Table \ref{table}, we plot the full DM evolution in the bottom panel of Figure \ref{fig:snr_revisit}. This demonstrates that DM is dominated by the SNR at early times ($t<t_{\rm ST}$) and then dominated by the ISM at late times ($t>t_{\rm ST}$). Furthermore, at intermediate times ($t\sim t_{\rm ST}$), the local DM contribution is actually rather constant. Also, note the power-law behavior at early and late times. At early times, it appears ${\rm DM}_{\rm local}\propto t^{-1/2}$, which is different than the scaling of ${\rm DM}\propto t^{-2}$ found from simple analytic arguments that assume a constant mass fraction of ionized material \citep[e.g.,][]{Connor16,Piro17,Metzger17,Yang17b}.

A wider range of DM solutions are summarized in Figure~\ref{fig:dm_constant}, where we consider a variety of $n_0$ values as well as $M=10\,M_\odot$ and $2\,M_\odot$ (note that we keep $E$ fixed at $10^{51}\,{\rm erg}$ for all these calculations). These masses are meant to represent the SNe that are from a red supergiant or a stripped-envelope progenitor (e.g., Type Ib/c), respectively. These solution highlight the fundamental role played by the timescale $t_{\rm ST}$, which is approximately the time at which $d{\rm DM}/{dt}$ switches from negative to positive. For the times with $t\sim t_{\rm ST}$, DM can be relatively constant for hundreds of years if not more. {\it Thus even if the DM of an FRB is not changing with time, this does not disprove the hypothesis of a rather young SN as the FRB progenitor.} At late times, the DM only depends on $n_0$, but interestingly, at early times both $M$ and $n_0$. This is different from simpler estimates of the ballistic phase which might assume that only $M$ is important at early times.

To better understand these simple scalings, and provide useful formulae for comparison to future observations, in the following sections we consider the behavior of the DM in the limits of early and late times.

\subsubsection{Constant Density: Ejecta-Dominated Stage DM Estimate}

Taking the limit $t\ll t_{\rm ST}$, and using the expressions given in Table \ref{table}, the thickness is roughly
\be
	R_c-R_r &\approx& 0.434 (t/t_{\rm ST})^{5/2}R_{\rm ST},
    \nonumber
    \\
    &\approx & 1.5\times10^{-6}
    E_{51}^{5/4}
    M_{1}^{-7/4}
    n_0^{1/2}
    t_{\rm yr}^{5/2}\,{\rm pc},
    \label{eq:r_c_r_r}
\ee
where $t_{\rm yr}=t/{\rm yr}$. In the limit of $t\ll t_{\rm ST}$, we also find $v_b\approx 1.37 v_{\rm ST}$ and $\tilde{v}_r\approx 1.41(t/t_{\rm ST})^{3/2}v_{\rm ST}$. Substituting these into Equations (\ref{eq:rho_r}) and (\ref{eq:ner}) then results in 
\be
	n_r = 3.77 n_0 \lp \frac{\mu}{\mu_e}\rp \lp \frac{t}{t_{\rm ST}}\rp^{-3}.
\ee
This demonstrates that the density is going down like $t^{-3}$, as one might assume for material expanding with constant velocity. Furthermore, since $t_{\rm ST}\propto n_0^{-1/3}$, this density is in fact independent of $n_0$ as one would expect during the ejecta-dominated phases.

\begin{figure}
\epsscale{1.2}
\plotone{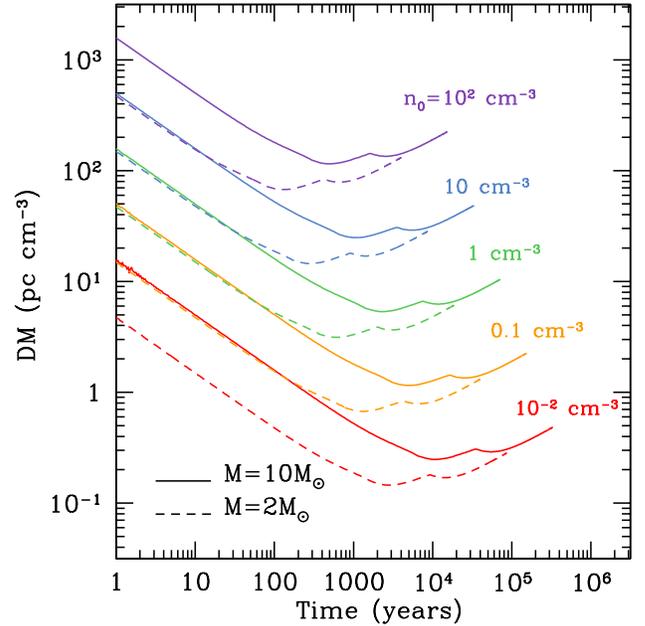}
\caption{{Time evolution of the DM for a red supergiant progenitor (solid lines) or stripped-envelope SN (e.g., Type Ib/c, dashed lines). These use $f=0$ and $E=10^{51}\,{\rm erg}$, with $M=10\,M_\odot$ or $M=2\,M_\odot$, respectively. Different colors lines indicate different values of $n_0$ as labeled.}}
\label{fig:dm_constant}
\epsscale{1.0}
\end{figure}

Using Equation (\ref{eq:dm_snr}), the dispersion measure of the SNR is given by
\be
	{\rm DM}_{\rm SNR} = 52.6 (\mu/\mu_e) E_{51}^{-1/4}
	M_1^{3/4}
	n_0^{1/2}
	t_{\rm yr}^{-1/2}\,{\rm pc\,cm^{-3}}.
	\nonumber
	\\
    \label{eq:DM_SNR}
\ee
As mentioned above, this scaling $\propto t^{-1/2}$ is very different from that found from previous simpler estimates that use $\propto t^{-2}$. The main difference is that those works assumed a constant fraction of material ionized. Instead, the ionized radial extent should scale with $R_c-R_r$, which is growing with time much faster than linearly as shown in Equation~(\ref{eq:r_c_r_r}). Another important difference is that this DM now includes a dependence on $n_0$ (as was seen in Figure \ref{fig:dm_constant}). This is because the larger the $n_0$ is, the more strongly the reverse shock is driven back into the ejecta to ionize the material.

As emphasized by \citet{Piro16} and \citet{Piro17}, an important discriminant for probing the environment of an FRB is how the DM potentially changes with time. Taking the time derivative of Equation (\ref{eq:DM_SNR}), we obtain
\be
	\frac{d{\rm DM}_{\rm SNR}}{dt} & =& -26.4 (\mu/\mu_e)E_{51}^{-1/4}
    \nonumber
    \\
	&&\times M_1^{3/4}
	n_0^{1/2}
	t_{\rm yr}^{-3/2}\,{\rm pc\,cm^{-3}\,yr^{-1}}.
\ee
Again, like DM$_{\rm SNR}$, we derive a shallower scaling with $t$ in comparison to other estimates in the literature due to a more realistic description of the ionized extent.

The other main region of free electrons is the shocked ISM material that is swept up within the region between the blastwave radius and the contact discontinuity,
\be
	R_b - R_c \approx 0.13v_{\rm ST}t.
	\label{eq:r_b_r_c}
\ee
Using Equation (\ref{eq:dm_ism}), this contribution is
\be
	{\rm DM}_{\rm ISM}
	\approx 5.4\times10^{-3}
	E_{51}^{1/2}
	M_1^{-1/2}
	n_0
	t_{\rm yr}\,{\rm pc\,cm^{-3}},
\ee
where we take $f=0$. For a neutral ambient medium (i.e., $f=0$), this DM is actually increasing with time as more and more ISM material is swept up. Nevertheless, the overall contribution is orders of magnitude smaller than the SN contribution (as seen on the lefthand side of Figure \ref{fig:snr_revisit}) and is not expected to be seen directly at early times.

\subsubsection{Constant Density: Sedov-Taylor Stage DM Estimate}

Next, in this limit $t\gg t_{\rm ST}$, the radial extent of the swept up ISM instead scales as
\be
	R_b-R_c &\approx& (1.56)^{2/5}(1-0.91)(t/t_{\rm ST})^{2/5}R_{\rm ST},
	\nonumber
	\\
	&=& 0.11(t/t_{\rm ST})^{2/5}R_{\rm ST}.
	\label{eq:r_b_r_c:ST}
\ee
The corresponding DM is
\be
	{\rm DM}_{\rm ISM} &=& 4n_0 (R_b-R_c) + f n_0 (R_{\rm ISM}-R_b),
	\nonumber
	\\
	&\approx& 1.8
	E_{51}^{1/5}
	n_0^{4/5}
	t_{1000\,\rm yr}^{2/5}\,{\rm pc\,cm^{-3}},
\ee
where $t_{1000\,\rm yr}=t/1000\,{\rm yr}$ and in the last expression we assume $f=0$. This is the same scaling as presented by \citet{Yang17b}, with a similar prefactor within $\approx~15\%$ of their result. Taking the derivative, we find
\be
	\frac{d{\rm DM}_{\rm ISM}}{dt} =
	 0.72 (1-2.7f)
	E_{51}^{1/5}
	n_0^{4/5}
	t_{1000\,\rm yr}^{-3/5}\,{\rm pc\,cm^{-3}},
    \nonumber
    \\
\ee
where here we have included a factor $-fn_0dR_b/dt$ due to ionized ISM material being swept by the forward shock. This shows that an {\em increasing}\ DM is possible for $f \la0.4$.

\subsection{Constant Density: Rotation Measure}

Shocks driven during the expansion of the SNR can generate magnetic fields that may imprint themselves on an FRB through Faraday rotation. Following \citet{Piro16}, we consider the magnetic fields generated by the forward and reverse shocks, assuming that the magnetic fields roughly obey equipartition with the shock velocities.

\begin{figure}
\epsscale{1.2}
\plotone{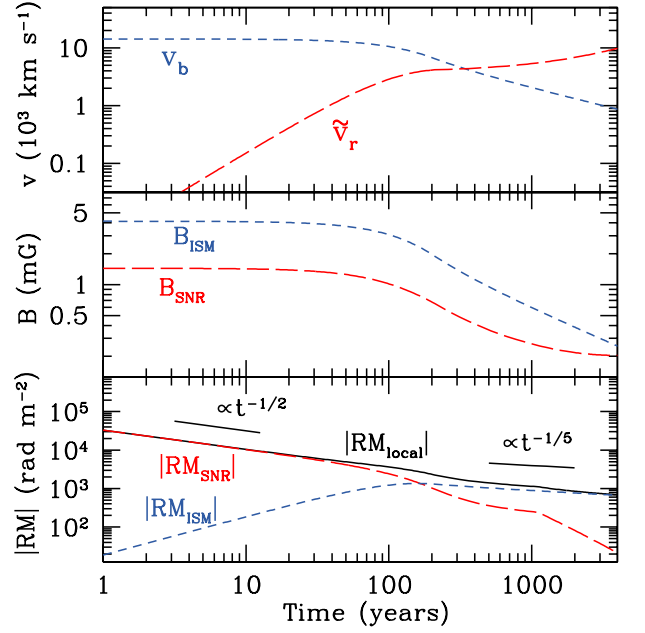}
\caption{Velocity, magnetic field, and RM evolution as a function of time for $\epsilon_B=0.1$, $M=M_\odot$, $E=10^{51}\,{\rm erg}$, and $n_0=1\,{\rm cm^{-3}}$. Red, long-dashed lines correspond to the reverse shock SNR features, while the blue, short-dashed lines correspond to the forward shock ISM contribution.}
\label{fig:bfield_rm}
\epsscale{1.0}
\end{figure}

For the reverse shock, the magnetic field is then
\be
	\frac{B_{\rm SNR}^2}{8\pi} \approx \epsilon_B \rho_r \tilde{v}_r^2/2 \Rightarrow B_{\rm SNR}\approx (4\pi \epsilon_B \rho_r)^{1/2}\tilde{v}_r,
    \label{eq:b_snr}
\ee
where $\epsilon_B$ is a parameter that sets how much of the shock energy goes into the magnetic field. Assuming equipartition between the forward shock and the magnetic field generated in the ISM, the corresponding field strength is
\be
	 B_{\rm ISM}\approx (16\pi \epsilon_B m_p n_0)^{1/2}v_b.
     \label{eq:b_ism}
\ee
The velocities and corresponding magnetic fields are plotted in the upper panel of Figure \ref{fig:bfield_rm} for $\epsilon_B=0.1$, $M=M_\odot$, $E=10^{51}\,{\rm erg}$, and $n_0=1\,{\rm cm^{-3}}$. This shows the general trend that the magnetic fields are rather constant at early times, but then decrease during the Sedov-Taylor phase.

\begin{figure}
\epsscale{1.2}
\plotone{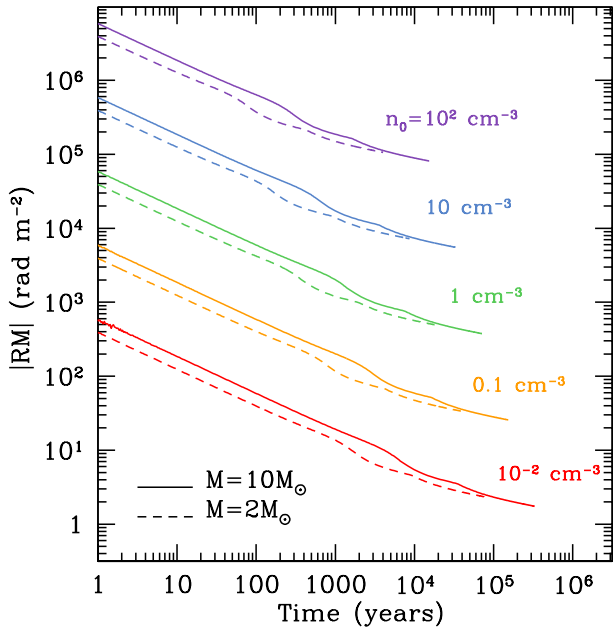}
\caption{Time evolution of the RM for $M=10\,M_\odot$ (solid lines) and $M=2\,M_\odot$ (dashed lines) for a red supergiant progenitor and stripped envelope SN, respectively, in both cases using $E=10^{51}\,{\rm erg}$. Different colors lines indicate different values of $n_0$ as labeled.}
\label{fig:rm_constant}
\epsscale{1.0}
\end{figure}

The associated rotation measure for a density of ionized $n$ with line-of-sight component of the magnetic field $B_{||}$ is 
\be
	{\rm RM}
    = \frac{e^3}{2\pi m_e^2 c^4} \int n B_{||} dl.
\ee
A useful relation for relating the RM and DM of the $i$-th region within the system is
\be
	{\rm RM}_i = 0.81 \lp\frac{{\rm DM}_i}{{\rm pc\,cm^{-3}}} \rp
    \lp\frac{B_i}{\mu {\rm G}} \rp {\rm rad\,m^{-2} }.
\ee
This expression is used to plot the RM evolution in the bottom panel of Figure \ref{fig:bfield_rm}. Here we assume that ${\rm RM}_{\rm SNR}$ and ${\rm RM}_{\rm ISM}$ can be simply added together to get ${\rm RM}_{\rm local}$. Just as for the DM evolution, the RM is dominated by the SNR at early times and the ISM at late times. The RM can be very large at early times, and the changes in RM can be quite substantial even if the changes in DM are small. Furthermore, while DM can be decreasing, roughly constant, or increasing depending on the time, the RM is strictly decreasing for this scenario. The full set of solutions for red supergiant and stripped envelope SNe are summarized in Figure \ref{fig:rm_constant}. Just as for the DM, at late times the RM only depends on $n_0$, while early on it depends on both $M$ and $n_0$. Unlike the DM, we do not include the ionized ISM material (highlighted with the ionized fraction $f$) since it is not clear that this material should have an ordered magnetic field. In the following sections, we derive the analytic scalings for these dependencies at both early and late times.

  \begin{deluxetable*}{lccc}
  \tablecolumns{10} \tablewidth{500pt}
 \tablecaption{Evolution for Wind Environment\tablenotemark{$a$}}
   \tablehead{  & Ejected-Dominated Stage $(t<t_{\rm ch})$ & Wind-Dominated Stage $(t>t_{\rm ch})$}
  \startdata
   Forward Shock &  $R_b/R_{\rm ch} = 1.79(t/t_{\rm ch})\lb 1+0.33(t/t_{\rm ch})^{1/2}\rb^{-2}$ & $R_b/R_{\rm ch} = \lb 1.11(t/t_{\rm ch})-0.11\rb^{2/3}$ \\
    & $v_b/v_{\rm ch} = 1.79\lb 1+0.33(t/t_{\rm ch})^{1/2} \rb^{-3}$  & $v_b/v_{\rm ch} = 0.74\lb 1.11(t/t_{\rm ch})-0.11\rb^{-1/3}$ \\
	Contact Discontinuity &  $R_c/R_{\rm ch} = 1.50(t/t_{\rm ch})\lb 1+0.33(t/t_{\rm ch})^{1/2}\rb^{-2}$ & $R_c/R_{\rm ch} = 0.84\lb 1.11(t/t_{\rm ch})-0.11\rb^{2/3}$ \\
    & $v_c/v_{\rm ch} = 1.50\lb 1+0.33(t/t_{\rm ch})^{1/2} \rb^{-3}$  & $v_c/v_{\rm ch} = 0.62\lb 1.11(t/t_{\rm ch})-0.11\rb^{-1/3}$ \\
   Reverse Shock & $R_r/R_{\rm ch} = 1.50(t/t_{\rm ch})\lb 1+0.70(t/t_{\rm ch})^{1/2} \rb^{-2}$ & $R_r/R_{\rm ch} =  (t/t_{\rm ch})\lb 0.54 - 0.02(t/t_{\rm ch}) -0.19\ln (t/t_{\rm ch}) \rb $\\
    & $\tilde{v}_r/v_{\rm ch}=  1.06(t/t_{\rm ch})^{1/2}\lb 1+0.70(t/t_{\rm ch})^{1/2} \rb^{-3}$ & $\tilde{v}_r/v_{\rm ch} = 0.19+0.02(t/t_{\rm ch})$\\
	DM (${\rm pc\,cm^{-3}}$)\tablenotemark{$b$} & $1.3\times10^4\mu_e^{-1} E_{51}^{-3/4}M_1^{5/4}K_{13}^{1/2}t_{\rm yr}^{-3/2}$ & $1.9\times10^{-2}\mu_e^{-1}E_{51}^{-1/3}K_{13}^{4/3}t_{10^4\,{\rm yr}}^{-2/3}$\\
	$|$RM$|$ (${\rm rad\,m^{-2}}$)\tablenotemark{$b$} & $2.0\times10^3 x_{0.1}(R_*/100\,R_\odot)
    (B_*/1\,{\rm G})\mu_e^{-1} E_{51}^{-1}M_1
    t_{\rm yr}^{-2}$ & $1.7\times10^{-3}x_{0.1} (R_*/100\,R_\odot)
    (B/1\,{\rm G})\mu_e^{-1} E_{51}^{-2/3}
    K_{13}^{5/3} t_{10^4\,{\rm yr}}^{-4/3}$
   \enddata
    \tablenotetext{$a$}{Analytic functions for shocks and contact discontinuity are derived in the Appendix. Also see Figure \ref{fig:snr_sln} for the plotting of these functions.}
    \tablenotetext{$b$}{These expressions for DM and RM are in the extreme limits of $t\ll t_{\rm ch}$ and $t\gg t_{\rm ch}$. For the more detailed evolution, one should consult \mbox{Figures \ref{fig:dm_wind} and \ref{fig:rm_wind}.}}
\label{tab:wind}
\end{deluxetable*}

\subsubsection{Constant Density: Ejecta-Dominated Stage RM Estimate}

From Equation (\ref{eq:rho_r}), $\rho_r^{1/2}\tilde{v}_r \approx (4n_0 \mu m_p)^{1/2} v_b$. Combining this with using $v_b\approx 1.37 v_{\rm ST}$ in the limit $t\ll t_{\rm ST}$, and substituting this into Equation (\ref{eq:b_snr}), the magnetic field is found to be roughly constant with time as
\be
	B_{\rm SNR} &\approx& 1.37(16\pi \epsilon_B \mu m_p n_0)^{1/2} v_{\rm ST},
	\nonumber
	\\
	 &\approx& 4.1\times10^{-3} \mu^{1/2}\epsilon_{-1}^{1/2}E_{51}^{1/2}M_1^{-1/2}n_0^{1/2}\,{\rm G},
    \label{eq:bfield}
\ee
where $\epsilon_{-1}=\epsilon_B/0.1$. The associated rotation measure is then
\be
	|{\rm RM_{SNR}}| &\approx&  1.8\times10^5(\mu^{3/2}/\mu_e) \epsilon_{-1}^{1/2}
    \nonumber
    \\
 	&&\times E_{51}^{1/4}  M_1^{1/4}n_0t_{\rm yr}^{-1/2}\,{\rm rad\,m^{-2}}.
\ee
This provides the $\propto t^{-1/2}$ scaling seen from the full solutions in Figures \ref{fig:bfield_rm} and \ref{fig:rm_constant}. Furthermore, we see directly that the RM depends on both $n_0$ and $M$.

\subsubsection{Constant Density: Sedov-Taylor Stage RM Estimates}

Using $v_b$ from Table \ref{table} for $t\gg t_{\rm ST}$ with Equation (\ref{eq:b_ism}),
\be
	B_{\rm ISM} \approx 5.6\times10^{-4} \epsilon_{-1}^{1/2}E_{51}^{1/5}n_0^{3/10}t_{1000\,\rm yr}^{-3/5}\,{\rm G}.
\ee
The corresponding RM is
\be
	|{\rm RM_{ISM}}| = 810 \epsilon_{-1}^{1/2}
		E_{51}^{2/5}  n_0^{11/10}t_{\rm 1000\,yr}^{-1/5}\,{\rm rad\,m^{-2}}.
\ee
The RM is indeed decreasing shallower than at early times and no longer depends on $M$.

\section{Wind Environment}
\label{sec:wind}

While the previous discussions assume a constant density ISM surrounding the SN, in many cases the circumstellar environment will be from a wind from the massive progenitor. This is likely especially important for FRBs if they come from young neutron stars \citep{Connor16,Piro16}. A wind can significantly alter the DM evolution, and also provide another source of magnetic field through the magnetized wind.

For a constant mass loss rate $\dot{M}$, we consider a constant velocity wind density profile
\be
	\rho_w = K/r^2,
\ee
where $K=\dot{M}/4\pi v_w$ and $v_w$ is the velocity of the wind. The wind mass loading parameter has a typical value of
\be
	K = 5.1\times10^{13}\dot{M}_{-5}v_6^{-1}\,{\rm g\,cm^{-1}},
\ee
where $\dot{M}_{-5}=10^{-5}M_\odot\,{\rm yr^{-1}}$ and $v_6 = v_w/10^6\,{\rm cm\,s^{-1}}$. Throughout our analysis we focus on varying $K$ rather than $\dot{M}$ and $v_w$ individually since this is the primary parameter that determines the evolution.

To better understand the SNR evolution under the influence of a wind environment, we derive a set of analytic equations for the characteristic radii in analogy to the constant ISM case. This derivation is provided in the Appendix, with a summary of the resulting analytic functions in Table \ref{tab:wind}. As with the constant density ISM case and the Sedov-Taylor scale, for the wind there is a characteristic radius and timescale which divides the ejecta-dominated and wind-dominated stages of the evolution. From the solutions in the Appendix these are found to be given by Equations (\ref{eq:R_ch}) and (\ref{eq:t_ch}), which when written in physical units are 
\be
	R_{\rm ch} = 16.8M_1K_{13}^{-1}\,{\rm pc},
\ee
and
\be
	t_{\rm ch} = 1.9\times10^3E_{51}^{-1/2}M_1^{3/2}K_{13}^{-1}\,{\rm yrs},
\ee
where $K_{13}=K/10^{13}\,{\rm g\,cm^{-1}}$.

\begin{figure}
\epsscale{1.2}
\plotone{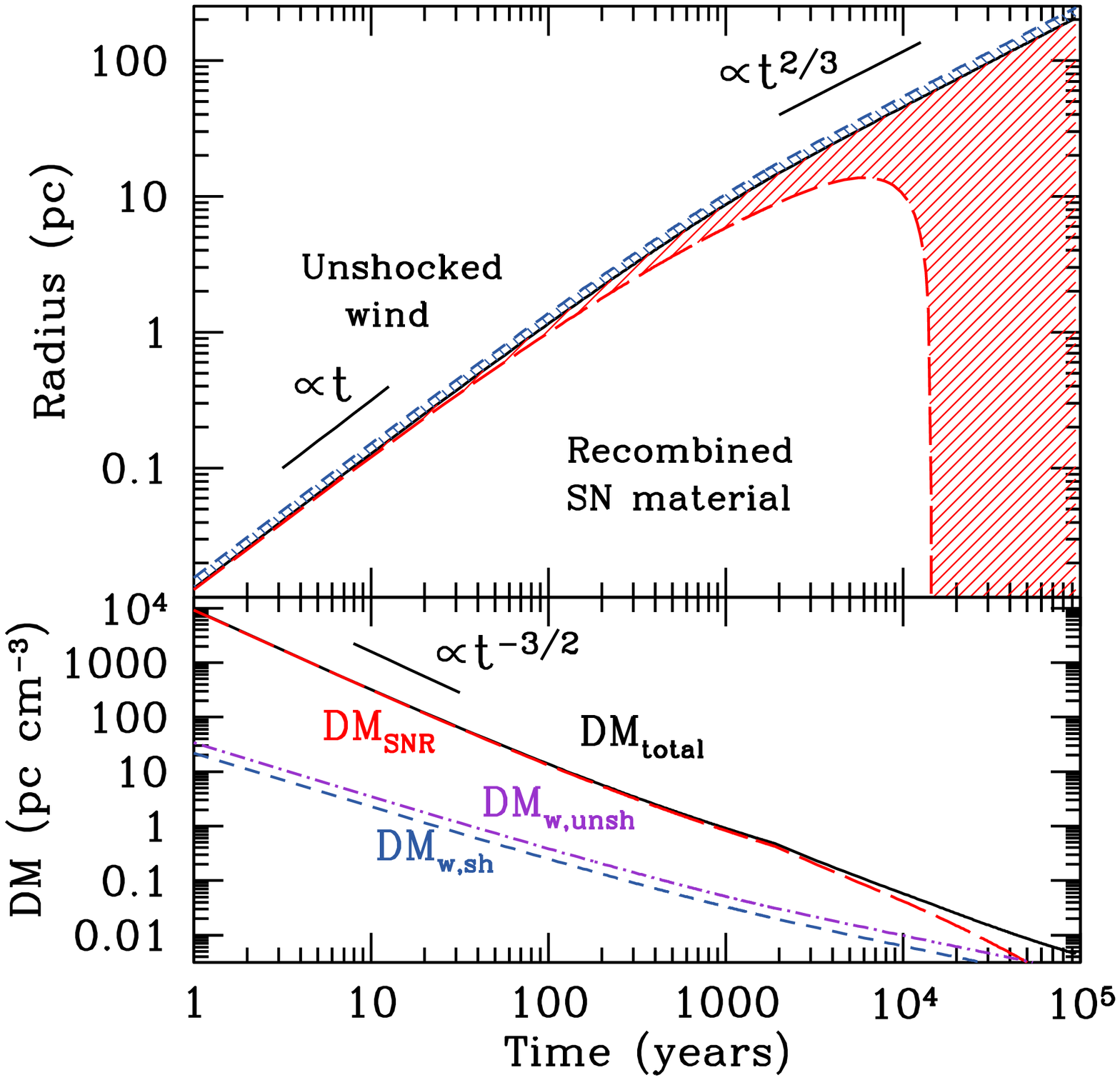}
\caption{Sample evolution of a SNR and the resulting DM for fiducial values $M=1,M_\odot$ and $E=10^{51}\,{\rm erg}$, expanding into a steady wind with $K=10^{13}\,{\rm g\,cm^{-1}}$; this combination corresponds to $t_{\rm ch} = 1.9\times10^3$~yr. The top panel shows the evolution of the three key radii $R_r$ (red long-dashed line), $R_c$ (black solid line), and $R_b$ (blue short-dashed line). The red and blue shaded regions denote the shocked SN ejecta and wind, respectively. The bottom panel shows how the DM evolves and is generally dominated by the SN ejecta, although if this were followed until even later times the wind would begin to contribute more}
\label{fig:snr_wind}
\epsscale{1.0}
\end{figure}

The general evolution of the SNR in the wind case is summarized in the upper panel of Figure \ref{fig:snr_wind}. This shows that in this case the blastwave evolves as
\be
  R_b\propto\begin{cases}
    t, & t\lesssim t_{\rm ch}\\
    t^{2/3}, & t\gtrsim t_{\rm ch},
  \end{cases}
\ee
which is steeper at late time in comparison to the constant ISM case. This is because the SNR is expanding into material that has a decreasing density with radius and thus not inhibited as strongly. This also means that the timescale $t_{\rm ch}$ can tend to be fairly long in comparison to the Sedov-Taylor timescale. For example, if we ask at what radius the wind density is similar to the constant density ISM, i.e., $\rho_w/(\mu_e m_p) = n_0$, we find
\be
	r = \lp \frac{K}{\mu_e m_p n_0} \rp^{1/2}
    = 0.79\mu_e^{-1/2} K_{13}^{1/2}n_0^{-1/2}\,{\rm pc},
    \label{eq:wind radius}
\ee
which is much less than $R_{\rm ch}$. This indicates that if  $t\gtrsim t_{\rm ch}$ is applicable to a given system, then the SNR is likely actually sitting within a bubble excavated by the wind.

\begin{figure}
\epsscale{1.2}
\plotone{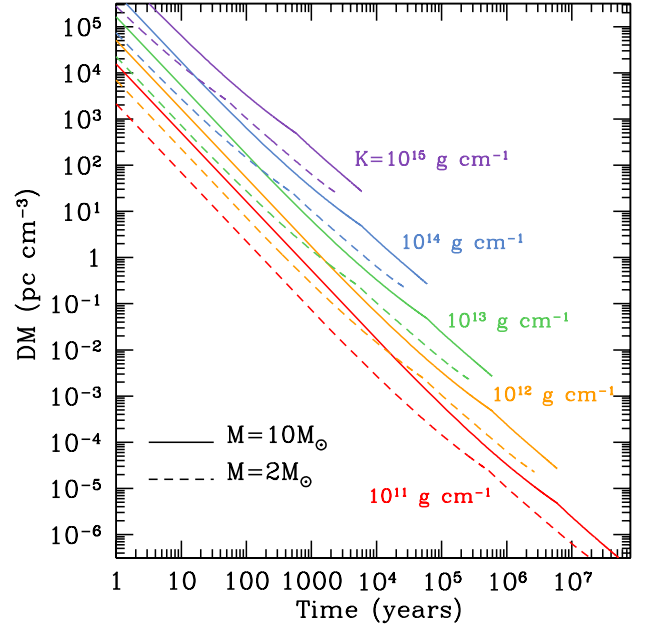}
\caption{Time evolution of the DM for a red supergiant progenitor (solid lines) or stripped-envelope SN (e.g., Type Ib/c, dashed lines). These use $E=10^{51}\,{\rm erg}$, with $M=10\,M_\odot$ or $M=2\,M_\odot$, respectively. Different colors lines indicate different values of $K$ as labeled.}
\label{fig:dm_wind}
\epsscale{1.0}
\end{figure}

\subsection{Wind: Dispersion Measure}

Similar to the constant density ISM case, we use pressure equality to solve for the electron density in the reverse shock region,
\be
	n_r \approx \frac{4\rho_w(R_b)}{\mu_e m_p} \lp \frac{v_b}{\tilde{v}_r}\rp^2.
    \label{eq:nr_wind}
\ee
where $\rho_w(R_b)=K/R_b^2$ is the density just ahead of the forward shock. From this we can again solve for the DM of the SNR using Equation (\ref{eq:dm_snr}).

{We assume in most cases that the wind itself will also have a significant ionized component, either because the wind is intrinsically ionized or because the shock breakout \citep{Matzner99} and subsequent shock cooling of the SN \citep{Nakar10,Piro13} will help ionize the wind. The wind's DM can then be broken into two components,} the shocked and un-shocked wind, which are determined according to
\be
	{\rm DM}_{w,\rm sh} = \frac{4\rho_w(R_b)}{\mu_em_p}(R_b-R_c),
\ee
and
\be
	{\rm DM}_{w,\rm unsh} &=& \int_{R_b}^\infty \frac{\rho_w}{\mu_e m_p} dr
    = \frac{K}{\mu_e m_p R_b},
\ee
respectively. The evolution of all the components ${\rm DM}_{\rm SNR}$, ${\rm DM}_{w,\rm sh}$, ${\rm DM}_{w,\rm unsh}$ are plotted in the bottom panel of Figure~\ref{fig:snr_wind}. Unlike the constant density case, here the DM is always strongly decreasing because even in the wind-dominated stage the wind density is getting smaller with radius. Over the timescales plotted here generally ${\rm DM}_{\rm SNR}\gg{\rm DM}_{w,\rm sh},{\rm DM}_{w,\rm unsh}$, even for $t>t_{\rm ch}$. We note though if this evolution were followed to even later timescale ($t\gtrsim10^6\,{\rm yrs}$ for these specific parameters) the wind component would begin to dominate.

Just as in the constant density case we next solve for the DM in the limits of early and late times.

\subsubsection{Wind: Ejecta-Dominated Stage DM Estimate}

Taking the limit $t\ll t_{\rm ch}$, the thickness of the region heated by the reverse shock is
\be
	R_c-R_r &=& 1.11 (t/t_{\rm ch})^{3/2} R_{\rm ch}
    \nonumber
    \\
    &=& 2.3\times10^{-4}
    E_{51}^{3/4} M_1^{-5/4}
    K_{13}^{1/2} t_{\rm yr}^{3/2}\,{\rm pc}.
\ee
This is generally found to be larger at early times than the constant density case because the large density near the star more readily pushed the reverse shock back into the ejecta. This though grows more slowly with time $\propto t^{3/2}$ rather than the constant density case that grows as $\propto t^{5/2}$.

To estimate the density of the reverse shocked region we use Equation (\ref{eq:nr_wind}) and approximate in the $t\ll t_{\rm ch}$ limit that $v_b\approx 1.78 v_{\rm ch}$ and $\tilde{v}_r\approx 1.16 (t/t_{\rm ch})^{1/2} v_{\rm ch}$ (from the relations in Table \ref{tab:wind}). This results in
\be
	n_r &\approx& 11.4\frac{\rho_w}{\mu_e m_p}\lp \frac{t}{t_{\rm ch}} \rp^{-1}
    \nonumber
    \\
    &\approx & 5.4\times10^7 \mu_e^{-1} E_{51}^{-3/2}M_1^{5/2}t_{\rm yr}^{-3}
    \,{\rm cm^{-3}},
\ee
Putting this together with the thickness of the shocked regions provides
\be
	{\rm DM}_{\rm SNR} = 1.3\times10^4 
    \mu_e^{-1} E_{51}^{-3/4}M_1^{5/4}
    K_{13}^{1/2}t_{\rm yr}^{-3/2}\,
    {\rm pc\,cm^{-3}}.
    \nonumber
    \\
\ee
This is much larger than the constant density case because of the extremely large density for the wind in close proximity to the SN, which is more effective for driving the reverse shock. The DM then falls off more quickly with time than the constant density case because of the decreasing density of the wind.

As noted above, the wind has two contributions to the DM, which are from the shocked and unshocked regions. The shocked wind has a thickness
\be
	R_b-R_c &=& 0.29(t/t_{\rm ch})R_{\rm ch}
    \nonumber
    \\
    &=& 2.6\times10^{-3}E^{1/2}_{51}M_1^{-1/2} t_{\rm yr}\,{\rm pc}.
\ee
The density of this region is estimated to just be the shocked wind density
\be
	\frac{4\rho_w}{\mu_e m_p}
    =1.0\times10^4\mu_e^{-1}E_{51}^{-1}M_1 t_{\rm yr}^{-2}\,{\rm cm^{-3}}.
\ee
Putting these together, the shocked wind contributes a dispersion measure of
\be
	{\rm DM}_{w,\rm sh} = 26.1\mu_e^{-1}E_{51}^{-1/2}M_1^{1/2}
    t_{\rm yr}^{-1}\,{\rm pc\,cm^{-3}}.
\ee
There is also a wind contribution from all of the unshocked wind material outside the radius of the forward shock
\be
	{\rm DM}_{w,\rm unsh} &=& \int_{R_b}^\infty \frac{\rho_w}{\mu_e m_p} dr
    = \frac{K}{\mu_e m_p R_b}
    \nonumber
    \\
    & = & 39.7\mu_e^{-1} E_{51}^{-1/2}
    M_1^{1/2}t_{\rm yr}^{-1}\,{\rm pc\,cm^{-3}}.
\ee
Since this scales the same as the shocked regions, these can just be added together to provide
\be
	{\rm DM}_{w,\rm tot} &=& {\rm DM}_{w,\rm sh}+{\rm DM}_{w,\rm unsh}
    \nonumber
    \\
    &=&
    65.8\mu_e^{-1} E_{51}^{-1/2}
    M_1^{1/2}t_{\rm yr}^{-1}\,{\rm pc\,cm^{-3}}.
\ee
Note that this is still subdominant to the SNR contribution at these times.

\subsubsection{Wind: Wind-Dominated Stage DM Estimate}

As mentioned above, the wind-dominated stage may only occur at very large times because $t_{\rm ch}$ is rather large. Nevertheless, with the caveat in mind, we can still solve for the DM. At sufficiently late times this is dominated by the shocked and unshocked wind material (even later than the times shown in Figure \ref{fig:snr_wind}).

In this late time limit, the width of the shocked wind material
\be
	R_b-R_c &=& 0.17(t/t_{\rm ch})^{2/3}R_{\rm ch}
    \nonumber
    \\
    &=& 8.64 E_{51}^{1/3}K_{13}^{-1/3}t_{10^4\,{\rm yr}}^{2/3}\,{\rm pc},
\ee
where $t_{10^4\,{\rm yr}}=t/10^4\,{\rm yr}$ and the density is
\be
	\frac{4\rho_w}{\mu_e m_p} = 8.5\times10^{-4}\mu_e^{-1}E_{51}^{-2/3}
    K_{13}^{5/3}t_{10^4\,{\rm yr}}^{-4/3}.
\ee
Putting these together results in a DM from the shocked region of
\be
	{\rm DM}_{w,\rm sh} = 7.3\times10^{-3}
    \mu_e^{-1}E_{51}^{-1/3}K_{13}^{4/3}t_{10^4\,{\rm yr}}^{-2/3}\,{\rm pc\,cm^{-3}}.
    \nonumber
    \\
\ee
Just as for early times, there is also a contribution from the unshocked wind as long as it is ionized. This is
\be
	{\rm DM}_{w,\rm unsh} &=&
    \frac{K}{\mu_e m_p R_b}
    \nonumber
    \\
    &=& 1.2\times10^{-2}\mu_e^{-1}E_{51}^{-1/3}K_{13}^{4/3}t_{10^4\,{\rm yr}}^{-2/3}\,{\rm pc\,cm^{-3}}.
    \nonumber
    \\
\ee
With the total DM being
\be
	{\rm DM}_{w,\rm tot} = 1.9\times10^{-2}\mu_e^{-1}E_{51}^{-1/3}K_{13}^{4/3}t_{10^4\,{\rm yr}}^{-2/3}\,{\rm pc\,cm^{-3}}.
    \nonumber
    \\
\ee
There is also a contribution from the SNR itself, but we ignore it here since it is comparable to the wind component we already account for and it does not have a simple power law solution. This is included in the plots though, such as Figures~\ref{fig:snr_wind} and \ref{fig:dm_wind}, and this is the reason there is still a non-negligible dependence on $M$ at the latest times plotted.

\begin{figure}
\epsscale{1.2}
\plotone{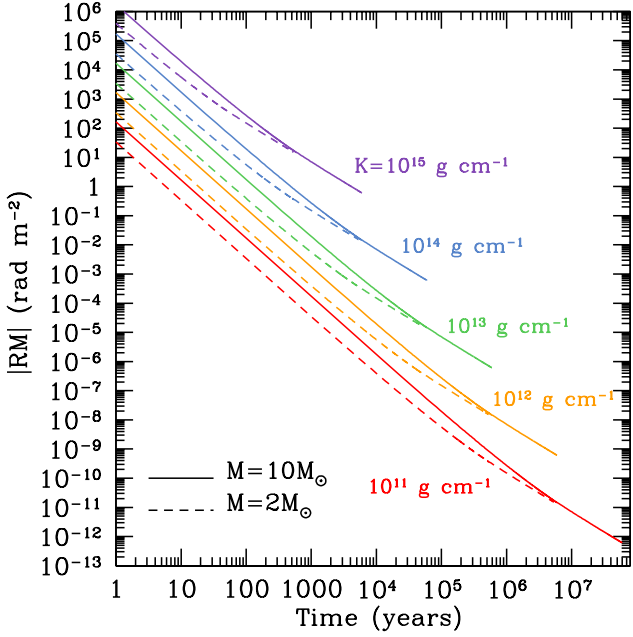}
\caption{Time evolution of the RM for $M=10\,M_\odot$ (solid lines) and $M=2\,M_\odot$ (dashed lines) for a red supergiant progenitor and stripped envelope SN, respectively, in both cases using $E=10^{51}\,{\rm erg}$, $x=0.1$, $B_*=1\,{\rm G}$, and $R_*=100\,R_\odot$. Different colors lines indicate different values of $K$ as labeled.}
\label{fig:rm_wind}
\epsscale{1.0}
\end{figure}

\subsection{Wind: Rotation Measure}

A wind environment is also interesting because it can provide an ordered magnetic field that can be swept up by the SNR. Thus for the wind case we focus on this possible contribution to the RM rather than shock generation of magnetic fields as for the constant density case.

Consider a toroidal magnetic field with the functional form
\be
	B_\phi(r) = B_* \lp \frac{v_{\rm rot}}{v_w}\rp \lp\frac{R_*}{r}\rp.
\ee
This is basically a split monopole that has been wrapped up by the star's rotation. The wind's contribution to the RM is determined by flux freezing of the swept up magnetic material \citep[as discussed by][]{Harvey-Smith10}. Once the forward shock has reached a radius $R_b$, the swept up magnetic field is
\be
	\int_{R_*}^{R_b} 2\pi r B_\phi (r) dr
    = 2\pi B_* x R_b R_*,
\ee
where $x \equiv v_{\rm rot}/v_w$ and we assume $R_b\gg R_*$. If the magnetic field within the shocked wind region is $B_\phi'$, then the magnetic flux of this material is
\be
	\int_{R_c}^{R_b}2\pi r B_\phi' dr =\pi (R_b^2-R_c^2)B_\phi'.
\ee
Equating these two fluxes allows us find
\be
	B_\phi' &= &2 B_*x \frac{R_bR_*}{R_b^2-R_c^2},
\ee
for the shocked magnetic field strength. The rotation measure is then given by
\be
	{\rm RM}_w
    = \frac{e^3}{2\pi m_e^2 c^4} \int_{R_c}^{R_b} \frac{4\rho_w(R_b)}{\mu_e m_p} B_\phi' dl.
\ee
This is plotted for a variety of different parameters in Figure~\ref{fig:rm_wind}. This demonstrates that the RM drops dramatically because of the combination of both the density and magnetic field strongly decreasing with time. Nevertheless, the RM can be very high at early times, especially if the magnetic field is larger than the modest field we assume here. Also note that $x$, $R_*$ and $B_*$ are fixed here even though in detail they should be different for different types of massive progenitors.

\subsection{Wind: Ejecta-Dominated Stage RM Estimate}

Using the expressions given above allow us to estimate the early-time magnetic field
\be
    B_\phi' = 96 x_{0.1}\lp\frac{R_*}{100\,R_\odot}\rp
    \lp \frac{B_*}{1\,{\rm G}}\rp E_{51}^{-1/2}M_1^{1/2}
    t_{\rm yr}^{-1}\,{\mu\rm G},
    \nonumber
    \\
    \label{eq:bphi}
\ee
where  $x_{0.1}=x/0.1$. The total RM is then
\be
	|{\rm RM}_w| &=& 2.0\times10^3 x_{0.1}\lp\frac{R_*}{100\,R_\odot}\rp
    \lp \frac{B_*}{1\,{\rm G}}\rp
    \nonumber
    \\
    &&\times\mu_e^{-1} E_{51}^{-1}M_1
    t_{\rm yr}^{-2}\,{\rm rad\,m^{-2}}.
\ee
Thus the RM contribution from the wind can be considerable.

\subsection{Wind: Wind-Dominated Stage RM Estimate}

For the late-time evolution, we use the same analytic expression from Equation (\ref{eq:bphi}) to derive
\be
	B_\phi' = 0.29x_{0.1}\lp\frac{R_*}{100\,{\rm G}}\rp
    \lp \frac{B}{1\,{\rm G}} \rp E_{51}^{-1/3}K_{13}^{1/3} t_{10^4\,{\rm yr}}^{-2/3}
    \,{\mu\rm G}.
    \nonumber
    \\
\ee
Multiplying this by the DM results
\be
	|{\rm RM}_w| &=& 1.7\times10^{-3}x_{0.1} \lp\frac{R_*}{100\,R_\odot}\rp
    \lp \frac{B}{1\,{\rm G}} \rp
    \nonumber
    \\
    &&\times\mu_e^{-1} E_{51}^{-2/3}
    K_{13}^{5/3} t_{10^4\,{\rm yr}}^{-4/3}\,{\rm rad\,m^{-2}}.
\ee
Thus the RM becomes somewhat shallower with time at late times, although it is so small at this point it may be negligible.

\section{Comparison to FRB Measurements}
\label{sec:discussion}

\begin{figure}
\epsscale{1.2}
\plotone{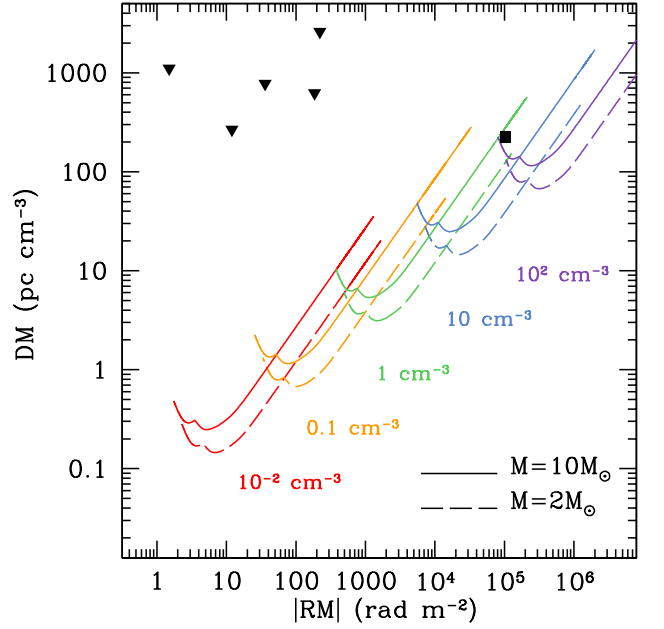}
\caption{The RM versus DM evolution for all the constant ISM density models considered in Section \ref{sec:constant}. In comparison, measured values for FRBs are shown with solid symbols. In the case of the repeating FRB 121102, the local DM can be known from the localization of the source, and thus this is plotted with a square. All other FRBs are plotted as upper limits on DM, since they are not localized and a significant fraction of their DM could be from the IGM.}
\label{fig:dm_rm_constant}
\epsscale{1.0}
\end{figure}

\begin{figure}
\epsscale{1.2}
\plotone{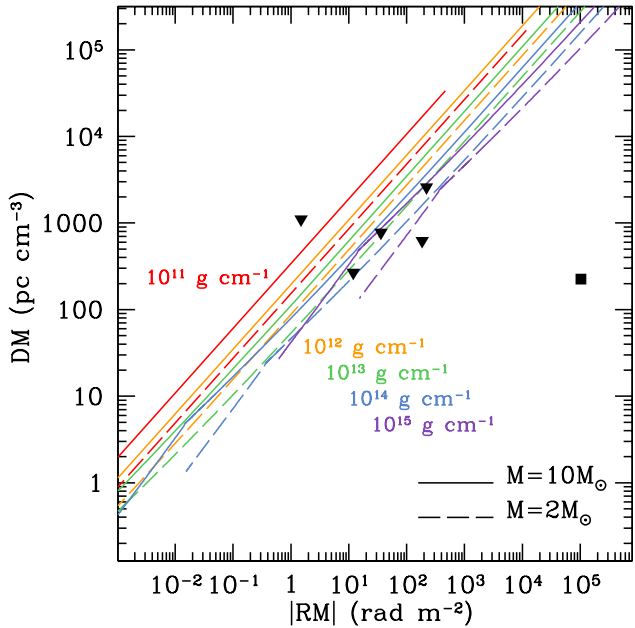}
\caption{The same as Figure \ref{fig:dm_rm_constant}, but for the wind models of Section \ref{sec:wind}.}
\label{fig:dm_rm_wind}
\epsscale{1.0}
\end{figure}

We now consider the implications of the DM and RM evolution described in the previous sections for the DMs and RMs observed for FRBs. In \mbox{Figures \ref{fig:dm_rm_constant} and \ref{fig:dm_rm_wind}}, we plot the RM versus DM evolution for all of the constant density and wind models considered in \mbox{Sections \ref{sec:constant} and \ref{sec:wind}.} As a comparison, we plot all FRBs with measured values of RM and DM with solid symbols. In the case of the repeating FRB 121102 \citep{Spitler14,Spitler16,Scholz16}, the local DM can be known from the localization of the source, and thus this is plotted with a square. Furthermore, its RM has been measured to vary from $(1.33-1.46)\times10^5\,{\rm rad\,m^{-2}}$ \citep{Michilli18}. All other FRBs are plotted as upper limits on DM, since they are not localized and a significant fraction of their DM could be from the IGM. {Their measured RMs are available in \citet{Masui15}, \citet{Petroff17}, \citet{Ravi16}, \citet{Keane16}, and \citet{Caleb18}, although we note in the case of FRB~150418 that the uncertainty is rather large with $|{\rm RM}|=36\pm52\,{\rm rad\,m^{-2}}$. In principle these RM values could instead be viewed as upper limits if there is additional magnetic fields in the IGM or host galaxy. Also, \citet{Caleb18} find an RM for FRB 151230 that is consistent with zero, and thus we do not include it in either \mbox{Figures \ref{fig:dm_rm_constant} or \ref{fig:dm_rm_wind}}.}

First examining Figure \ref{fig:dm_rm_constant}, we see that FRB~121102, which has the best known values for these properties, is actually fairly consistent with these estimates if the ISM is sufficiently dense ($n_0\sim100\,{\rm cm^{-3}}$). Furthermore, a large $n_0$ would help the $d{\rm DM}/dt$ to be rather small as been observed for this FRB over many years because $t\sim t_{\rm ST}$. This would imply an age of the SNR of $\sim10^2-10^3\,{\rm yrs}$, depending on the mass of the ejecta, which would still be a young NS, but old enough that free-free absorption of the FRB should not be a problem {as described in the theoretical work of \citet{Piro16} or the empirical study by \citet{Bietenholz17}. Most recently, it has been revealed that the RM of FRB~121102 has decreased over a $\sim$7 month timescale, while the DM has remained relatively constant \citep{Michilli18}. This is again qualitatively consistent with our results when the SNR is near the Sedov-Taylor timescale.} The other FRBs are potentially more difficult to reconcile with this picture. Although the DM values are upper limits, the low RM values indicate that the local DM must be very small. Furthermore, if this is the case, then it would be difficult to satisfy both the DM and RM unless the ISM densities are much smaller than what we infer for FRB~121102.

Comparing to Figure \ref{fig:dm_rm_wind}, the situation is seemingly reversed. Now it is FRB~121102 that is inconsistent with any of the models unless the magnetic field were a factor of $\sim10^4$ higher. On the other hand, the other FRBs are fairly consistent with the wind models. Even though these DM values are upper limits, they could still be reconciled if lower by a factor of $\sim10$ or more by just adjusting the magnetic field of the progenitor star.

An outstanding question remains of whether all FRBs are the same or if the repeater should be considered in a separate class. Interestingly, these comparisons here argue that the combined DM-RM values are yet another way the repeater FRB~121102 appears to be unique compared to the other FRBs. This may mean that the environments are fundamentally different. Instead though, it could be that the environments are actually similar, but that the repeater is being observed in a different stage of evolution. As Equation~(\ref{eq:wind radius}) highlights, the wind may not extend as far as the typical Sedov-Taylor length scale. Thus, one could imagine that a given system could be wind dominated at early times (like the non-repeaters appear to be) but be more like a constant density ISM case at later times (like the repeater). Comparisons like this will be important in the future to better classify the ways in which FRBs are different or the same.

%\begin{itemize}
%\item Need to include redshift corrections
%\item Magnitude of contribution
%\item Prospects for detecting time evolution
%\item Likely wind properties of a low-metallicity progenitor as seen for the potential host of the repeating FRB
%\end{itemize}

\section{Conclusions and Discussion}
\label{sec:conclusions}

Motivated by the hypothesis that FRBs are from young neutron stars and thus should be embedded within SNRs, we have revisited the impact of an SNR on FRBs. This includes both constant density ISM and wind environments, and for the latter case we derived new analytic solutions for the SNR evolution summarized in Appendix \ref{sec:appendix} and Table \ref{tab:wind}. In each case, we provided analytic expressions both for the DM and RM values. These are split into early times, which correspond to the stage when the blastwave is moving at constant velocity ($t<t_{\rm ST}$ or $t<t_{\rm ch}$ for the constant density ISM and wind cases, respectively) and late times, which is when the SNR has swept up an amount of material comparable to its mass ($t>t_{\rm ST}$ or $t>t_{\rm ch}$). Our main conclusions are as follows.
\begin{itemize}
\item The DM and RM are mostly determined by two regions: SN ejecta heated by the reverse shock and the surrounding material heated by the forward shock.
\item At early times, the DM is dominated by the SN ejecta, but it is {\it not} the case that ${\rm DM}\propto t^{-2}$ as normally assumed in the literature. This is because of the dynamics of the reverse shock, which results in a shallower scaling for DM and a dependence on the density of the surrounding medium.
\item At intermediate times ($t\sim t_{\rm ST}$), the DM for the constant density ISM case can be rather constant for hundreds of years if not more, so that a young neutron star hypothesis should not be ruled out if DM is not observed to change for a repeating FRB. On the other hand, the RM is found to always be decreasing.
\item {For the wind case, the DM always decreases with time. Furthermore, a magnetized wind swept up by the SN provides another region that may contribute to the RM observed for FRBs.}
\item The DM and RM for the repeating FRB~121102 appear consistent with the constant density case if the ISM density is large ($n_0\sim100\,{\rm cm^{-3}}$), which would also help explain why $d{\rm DM}_{\rm SNR}/dt$ is small. This implies an age of the FRB progenitor of $\sim10^2-10^3\,{\rm yrs}$, depending on the SN ejecta mass. {Furthermore, its decreasing RM while the DM is relatively constant is again qualitatively consistent with this interpretation.}
\item A constant density ISM is difficult to reconcile with the other FRBs (because of their lower RM values) unless a significant fraction \mbox{($>99.9\%$)} of their DM is from the IGM and host galaxy.
\item On the other hand, the wind case seems to naturally fit with most FRBs that are not the repeater. If this explains their DMs and RMs, it would argue that these FRBs are rather young and thus should have strongly decreasing DM and RM values if seen to repeat.
\item A significant contribution can be made to the RM even if the DM is not dominated by the SNR and is instead mostly due to the IGM. This means one should be cautious about inferring a magnetic field from observations by using the ratio RM/DM \citep[as done in][]{Ravi16} if different regions of electrons are contributing to each of these quantities.
\end{itemize}
{Considering the final point, the magnetic field generating the RM may be estimated when the RM and/or DM vary, since this helps separate the contribution of free electrons near to the FRB from the IGM contribution. For example, \citet{Katz18} shows that using the upper bound on the variation in DM when RM varies can place a lower bound on the magnetic field.}

We emphasize though that simply assuming a given system will only be the constant density case or wind case is probably an over simplification. In general, one could imagine a SNR at first mostly being dominated by a wind, but then evolving to a constant density case once it has overtaken the extent of the wind. In such cases, as highlighted by the discussion of the wind extent at the beginning of Section \ref{sec:wind} and Equation (\ref{eq:wind radius}), one might expect the $t<t_{\rm ch}$ solutions to be most applicable at early times, but actually the $t>t_{\rm ST}$ solutions to apply later.

This issue, as well as our currently simplistic treatment for following the contact discontinuity (see the discussion at the beginning of Section \ref{sec:constant}), argue that the next stage for this research necessitates numerical models of the SNR evolution. This would allow for more complicated density distributions for the surrounding material. In addition, it would allow us to consider a more realistic density distribution for the SNR itself, where instead of just assuming a constant density sphere as done here, it should in fact have a steep outer density gradient \citep[e.g.,][]{Truelove99}. {Looking beyond this, multi-dimensional simulations would be useful to resolve the complicated filamentary density that is seen for real SNRs. This may cause the DM and RM to vary significantly from what we calculate here, and thus our work represents the average properties at any given time. Such simulations would help for understanding the size and statistical properties of the deviations from this average.}

Ultimately though, one would like to see more repeating FRBs, since this work demonstrates that changes in the DM and RM values can strongly constrain the environment of the FRB. Even in the comparisons shown in Figures \ref{fig:dm_rm_constant} and \ref{fig:dm_rm_wind} there appears to be some dichotomy between the repeater and those FRBs that have not been seen to repeat. Actually localizing some of these other FRBs that have both a DM and RM measurement would allow the IGM component of their DMs to be subtracted. This would improve our understanding of their local DMs, and we would have a better idea of how different these bursts really are. In lieu of this, large statistical samples of FRBs may also be helpful, as expected by the Canadian Hydrogen Intensity Mapping Experiment \citep[CHIME;][]{chimefrb}. CHIME will be especially important because its low frequency range of $400-800\,{\rm MHZ}$ is sensitive to the free-free absorption cutoff expected from SNRs \citep{Piro16}, providing additional information about the age of the system that can be folded into the analysis presented here.

\acknowledgments

A.L.P. acknowledges partial support from the Research Corporation for Scientific Advancement (RCSA) for participation in the meeting Fast Radio Bursts: New Probes of Fundamental Physics and Cosmology at the Aspen Center for Physics (February 12--17, 2017) where the original seeds for this work were inspired. B.M.G. acknowledges the support of the Natural Sciences and Engineering Research Council of Canada (NSERC) through grant RGPIN-2015-05948, and of the Canada Research Chairs program.
The Dunlap Institute is funded through an endowment established by the David Dunlap family and the University of Toronto. The Aspen Center for Physics is supported by National Science Foundation grant PHY-1066293.

\begin{appendix}
\counterwithin{figure}{section}

\section{Supernova Remnant Evolution with Wind}
\label{sec:appendix}

Here we derive the evolution equations for a SNR surrounded by a constant velocity wind. For a mass loss rate of $\dot{M}$ we consider a density profile
\be
	\rho_w = K/r^2,
\ee
where $K=\dot{M}/4\pi v_w$ and $v_w$ is the velocity of the wind. Just as with the constant density ISM case, there are characteristic scales in this case analogous to the Sedov-Taylor scales. Here we just refer to these with the subscript ``ch'' for characteristic, and from dimensional analysis the characteristic radius and timescale must obey
\be
	R_{\rm ch} \propto MK^{-1},
    \label{eq:r_ch_prop}
\ee
and
\be
	t_{\rm ch} \propto E^{-1/2}M^{3/2}K^{-1},
    \label{eq:t_ch_prop}
\ee
respectively. Also useful is the relation between the SN energy and the maximum ejecta velocity
\be
	E = (3/10)Mv_e^2,
\ee
which we will be using throughout the derivation.

\subsection{Ejecta-Dominated Stage, $t<t_{\rm ch}$}

Just as for the constant density case, the evolution can be separated into two stages. These are an ejecta-dominated stage for $t< t_{\rm ch}$ and a Sedov-Taylor stage for $t>t_{\rm ch}$. We derive the general evolution in each stage and then require continuity to connect the two solutions. Here we start with the ejecta-dominated stage.

As shown in Figure \ref{fig:diagram}, we envision mass $M$ ejected in a SN explosion, which generates a contact discontinuity $R_c$ with corresponding forward shock and reverse shocks with radii $R_b$ and $R_r$, respectively, as it moves into the surrounding medium. A key property of the SNR is the pressure ratio between the forward and reverse shocks,
\be
	\phi(t) \equiv \frac{\rho_e(t) \tilde{v}_r^2(t)}{\rho_w(v_et)v_b^2(t)},
    \label{eq:phi}
\ee
where $\rho_e(t) = 3M/4\pi v_e^3t^3$ is the density of the ejecta, $\rho_w(v_et)=K/(v_et)^2$ is the wind density at a radius $v_et$, $v_b = dR_b/dt$ is the blastwave (forward shock) velocity, and
\be
	\tilde{v}_r \equiv \frac{R_r}{t} - v_r = \frac{R_r}{t} - \frac{dR_r}{dt},
    \label{eq:tilde_vr}
\ee
is the velocity of the reverse shock in the rest frame of the unshocked ejecta just ahead of it (defined to be positive).

Following \citet{McKee95}, a key principle we will use for finding analytic solutions to the evolution is assuming that this pressure ratio is roughly constant and equal to the value found in numerical simulations of the ejecta-dominated stage
\be
	\lim_{t\rightarrow0} \phi(t) \equiv \phi_{\rm ED} \approx 0.212,
\ee
where this specific value is from the numerical calculations by \citet{Truelove99}. The other key estimate is the ratio of the blastwave to the contact discontinuity, $\ell=R_b/R_c$, also known as the lead factor. Again we assume that this is constant and approximated by the ejected-dominate stage
\be
	\lim_{t\rightarrow0} \ell \equiv \ell_{\rm ED} \approx 1.19,
\ee
where this value is taken from the work of \citet{Hamilton84}. As with $\phi(t)$, we take the early time limit where $R_r\approx R_c$ and thus also approximate $R_b \approx \ell_{\rm ED}R_r$ and $v_b \approx \ell_{\rm ED}v_r$ for $t\rightarrow 0$.

As an aside, one could instead use mass conservation and assume a constant density behind the forward shock to estimate the lead factor. From mass conservation, comparing the shocked mass to the swept up mass, one finds
\be
	\frac{4\pi}{3}(R_b^3-R_c^3) 4\rho_w = \frac{4\pi}{3}R_b^3 \rho_w, 
\ee
where the factor of 4 is from the compression at the forward shock. Solving this equation leads to $R_b = (4/3)^{1/3}R_c$ or a lead factor of $\ell =(4/3)^{1/3}\approx 1.10$, slightly smaller than the value $\ell_{\rm ED}$ we use above. This is because in reality the density is not exactly constant in the region between the forward shock and the contact discontinuity.

Using these above approximations, we can then simplify Equation (\ref{eq:phi}) to the form
\be
	\tilde{v}_r = C_{\rm ED} \phi_{\rm ED}^{1/2} t^{1/2}v_r,
    \label{eq:tilde_vr2}
\ee
where we have introduced the constant
\be
	C_{\rm ED} \equiv \ell_{\rm ED}\left( \frac{4\pi v_e K}{3M} \right)^{1/2}.
\ee
We can alternatively write Equation (\ref{eq:tilde_vr2}) in terms of $R_r$,
\be
	\frac{R_r}{t} - \frac{dR_r}{dt} = C_{\rm ED} \phi_{\rm ED}^{1/2} t^{1/2}\frac{dR_r}{dt},
\ee
which is a first-order differential equation in $R_r$. Integrating this equation with the requirement that $R_r(t) \approx v_et$ for $t\rightarrow 0$ results in
\be
	R_r(t) = v_e t \left(1 + C_{\rm ED} \phi_{\rm ED}^{1/2}t^{1/2} \right)^{-2}.
\ee
Utilizing Equation (\ref{eq:tilde_vr}),
\be
	\tilde{v}_r(t) = v_e C_{\rm ED} \phi_{\rm ED}^{1/2}t^{1/2}
    		\left(1 + C_{\rm ED} \phi_{\rm ED}^{1/2}t^{1/2} \right)^{-3}.
\ee
Again matching the early-time limits, the blastwave radius and velocity are given by
\be
	R_b(t) = \ell_{\rm ED}v_e t \left(1 + C_{\rm ED} \phi_{\rm eff}^{1/2}t^{1/2} \right)^{-2},
    \label{eq:rb_ed}
\ee
and
\be
	v_b(t) = \ell_{\rm ED}v_e \left(1 + C_{\rm ED} \phi_{\rm eff}^{1/2}t^{1/2} \right)^{-3},
    \label{eq:vb_ed}
\ee
where we have replaced $\phi_{\rm ED}$ with $\phi_{\rm eff}\lesssim \phi_{\rm ED}$ to represent the loss of pressure felt by the forward shock as the SNR evolves away from the ejected-dominated stage (which has a stronger effect on the forward shock in comparison to the reverse shock). As we shall show below, the continuity conditions allow us to uniquely calculate $\phi_{\rm eff}$.

\subsection{Wind-Dominated Stage, $t>t_{\rm ch}$}

For sufficiently large $t$, the SNR evolution must obey the classical Sedov-Taylor solution for the wind profile \citep{Ostriker88}, which is given by
\be
	R_b(t) = \lp \frac{3}{2\pi} \frac{E}{K}\rp^{1/3} t^{2/3} = C_{\rm ch}t^{2/3}.
\ee
Taking the derivative of this expression
\be
	\frac{dR_b}{dt} = \frac{2}{3}C_{\rm ch}^{3/2} R_b^{-1/2},
    \label{eq:st_derivative}
\ee
Integrating this with the boundary condition that $R_b(t_{\rm ch})=R_{\rm ch}$ results in
\be
	R_b(t) = \left[ R_{\rm ch}^{3/2} + C_{\rm ch}^{3/2}(t-t_{\rm ch})\right]^{2/3},
\ee
for the general form of the blastwave radius.

The reverse shock is only weakly accelerated during the Sedov-Taylor phase, as represented by the small factor of $0.03$ in the expressions for $R_r$ and $\tilde{v}_r$ in the Sedov-Taylor stage in Table \ref{table}. The exact value can be calibrated with numerical simulations, but here we just assume, similar to the constant density case, a small acceleration with $\tilde{a}_r\approx 0.1 \tilde{v}_r(t_{\rm ch})/t_{\rm ch}$. The exact value of this does not impact our DM calculations since for $t>t_{\rm ch}$ the DM is dominated by swept up wind material. Integration with constant acceleration then gives
\be
	\tilde{v}_r = \tilde{v}_r(t_{\rm ch}) + \tilde{a}_r(t-t_{\rm ch}).
\ee
Next, we solve the differential Equation (\ref{eq:tilde_vr}) to find $R_b(t)$. This is facilitated by making a change of variables $u=R_r/t$, using the fact that $\tilde{v}_r = -tdu/dt$, solving for $u(t)$, and then transforming back to $R_r(t)$, resulting in
\be
	R_r(t) = t
   \left\{
   		R_r(t_{\rm ch})/t_{\rm ch} - \tilde{a}_r(t-t_{\rm ch})
   		- \left[ \tilde{v}_r(t_{\rm ch}) - \tilde{a}_rt_{\rm ch}\right]
   		\ln(t/t_{\rm ch})
    \right\},
\ee
for the reverse shock evolution.

\subsection{Connecting the Stages}

\begin{figure}
\epsscale{0.55}
\plotone{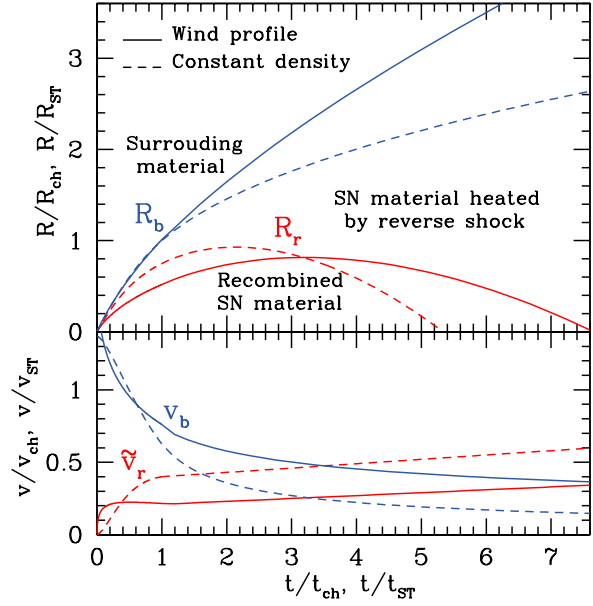}
\caption{Comparison of the evolution with a wind density profile (solid lines) to the case of a constant density ISM (dashed lines). The wind case is plotted in units of the characteristic properties, $R_{\rm ch}$ and $v_{\rm ch}$, while the constant density case is plotted in units of $R_{\rm ST}$ and $v_{\rm ST}$. The analytic expressions for each of these curves are summarized in \mbox{Tables \ref{table} and \ref{tab:wind}.}}
\label{fig:snr_sln}
\epsscale{1.0}
\end{figure}

Exact expressions for $t_{\rm ch}$ and $R_{\rm ch}$ can be derived by requiring continuity of solutions between the ejecta-dominated and wind-dominated stages.
Utilizing Equation (\ref{eq:rb_ed}), (\ref{eq:vb_ed}), and (\ref{eq:st_derivative}) and requiring continuity of $R_b$ and $v_b$ results in the expressions,
\be
	\ell_{\rm ED} v_e t_{\rm ch}  \left( 1+C_{\rm ED}\phi_{\rm eff}^{1/2}  t_{\rm ch}^{1/2}\right)^{-2} = R_{\rm ch},
    \label{eq:equation1}
\ee
and
\be
	\ell_{\rm ED} v_e  \left( 1+C_{\rm ED}\phi_{\rm eff}^{1/2}  t_{\rm ch}^{1/2}\right)^{-3}
    = \frac{2}{3} C_{\rm ch}^{3/2}R_{\rm ch}^{-1/2}.
    \label{eq:equation2}
\ee
These coupled equations have two unknowns $R_{\rm ch}$ and $t_{\rm ch}$ that can be solved for algebraically. Since we know the scaling expected for $R_{\rm ch}$ and $t_{\rm ch}$ from Equations (\ref{eq:r_ch_prop}) and (\ref{eq:t_ch_prop}), this process is easiest if we substitute
\be
	R_{\rm ch}=AMK^{-1},
\ee
and
\be
	t_{\rm ch}=BE^{-1/2}M^{3/2}K^{-1},
\ee
where $A$ and $B$ are dimensionless. This allows all dimensional factors to cancel from Equations (\ref{eq:equation1}) and (\ref{eq:equation2}). Combining the two equations allows us to cancel $B$ and find a family of solutions $\phi_{\rm eff}(A)$. A critical point is calculated from this function, defined as when $d\phi_{\rm eff}(A)/dA = 0$, which results in a value of $\phi_{\rm eff}\approx 0.0479$. This can then be substituted back in to find $A$ \mbox{and $B$.}

The two characteristic scales are then
\be
	R_{\rm ch} = 0.26 MK^{-1},
    \label{eq:R_ch}
\ee
and
\be
	t_{\rm ch} = 0.21 E^{-1/2}M^{3/2}K^{-1}.
    \label{eq:t_ch}
\ee
With these values, the other important constants become
\be
	C_{\rm ED} = \ell_{\rm ED}\left(\frac{4\pi}{3}\right)^{1/2} \left(\frac{10}{3}\right)^{1/4}
    E^{1/4} M^{-3/4}K^{1/2}
    = 1.53t_{\rm ch}^{-1/2}
\ee
and
\be
	C_{\rm ch} = \left( \frac{3}{2\pi}\right)^{1/3} E^{1/3} K^{-1/3}
    = 1.07t_{\rm ch}^{-2/3} R_{\rm ch}.
\ee
These can all be substituted back into the time evolution equations summarized above to solve for $R_b(t)$, $v_b(t)$, $R_r(t)$, and $\tilde{v}_r(t)$ in both the ejecta-dominated and wind-dominated stages. These results are summarized in Table \ref{tab:wind}.

A comparison of the solutions found here to the case of a constant density ISM are plotted in Figure \ref{fig:snr_sln}. During the ejecta-dominated stage, both cases show similar evolution for $R_b$, but the wind case shows stronger evolution of $R_r$ because the early high densities push the reverse shock back into the ejecta more strongly. At later times, the wind case is more gradual, since the blastwave is moving into lower density material. This causes the reverse shock to finally reach the center of the SN ejecta at later times as well.

\end{appendix}

\end{document}